# An Exactly Solvable Model of an Avalanche-Type Measuring Device: Macroscopic Distinctiveness and Wavefunction Collapse

R. Merlin

*The Harrison M. Randall Laboratory of Physics, The University of Michigan,
Ann Arbor, MI 48109-1040*

A quantum many-body model is presented with features similar to those of certain particle detectors. The energy spectrum contains a single metastable 'ready'-state and macroscopically-distinct 'pointer' states. Measurements do not pose paradoxes or require interventions outside the field theory formalism. Transitions into classical-like states can be triggered by a single particle with help of the thermal bath. Schrödinger cat states are associated with superpositions of inequivalent vacua, thus relating wavefunction collapse to the dynamics of symmetry breaking in phase transformations.

Nearly three-quarters of a century after the foundations of quantum mechanics were laid, the measurement problem and its associated inconsistencies, as typified by the notorious Einstein-Podolski-Rosen [1] and Schrödinger cat [2] paradoxes, remain largely unexplained [3,4,5,6]. Consider a measuring device which, following the interaction with a particle, evolves into state $|P_1\rangle$ if the particle was originally in state $\chi_1$ or into $|P_2\rangle$ if the particle state was $\chi_2$. Because the quantum evolution operator is strictly linear, if the initial state of the particle is $\chi = c_1\chi_1 + c_2\chi_2$, the final state of the combined system becomes

$$c_1\chi_1 \otimes |P_1\rangle + c_2\chi_2 \otimes |P_2\rangle \qquad (1)$$

which represents a linear superposition of macroscopically distinct states. Since such entangled states are at odds with reality, for the result must be either $|P_1\rangle$ or $|P_2\rangle$, it is usually assumed that the description of the measurement process lies outside quantum mechanics. In a broad sense, most physicists nowadays adhere to various variants of the Copenhagen interpretation, particularly to the so-called standard model which rests on, both, von Neumann's projection postulate (wavefunction collapse) and Born's statistical postulate relating the outcome of a measurement to the square of the absolute value of the wavefunction. An interesting approach to solving the measurement problem is to invoke the phenomenon of decoherence, namely, the fact that interactions with the environment quickly remove interfering effects between macroscopically distinct states [7]. Without a mathematical description of the apparatus [8], however, it is difficult to assess the precise role decoherence plays in an actual measurement and, moreover, it is apparent that decoherence is not a substitute for and does not explain wavefunction collapse [4,9].

In this Letter, we introduce a quantum mechanical model whose dynamics closely resembles that of chemical reactions and detectors such as Glaser's bubble chamber, the Geiger counter and the AgBr crystal of the photographic plate, for which a single elementary particle can trigger a chain-like reaction and, thus, induce a macroscopic change. As shown pictorially in Fig. 1, our device involves two fictitious particles, *A* and *B*, both of which are bosons. Those of the *B*-type are the particles being detected while the *A*-ones are the 'stuff' the pre-measurement state of the detector is made of. The *B* particles occur in two forms, $B_\xi$ and $B_\eta$, which can be discriminated by the detector. Our model for the device can be solved exactly. Its main features are the following:

(*i*) The spectrum of the isolated detector exhibits a high-energy 'ready'-state $|M\rangle$, composed only with bosons of the *A*-type, and two separate continua ending, on the low-energy side, in the 'pointer' states $|G_\xi\rangle$ and $|G_\eta\rangle$, which contain a macroscopic number of, respectively, $B_\xi$ and $B_\eta$ particles. Hence, the $|M\rangle$ state and the two pointer states are macroscopically distinct in Leggett's sense [4].

(*ii*) The device is a *B*–detector. With the crucial assistance of the thermal bath, a single particle can initiate a chain-like reaction process which begins at $b^+_{\xi,\eta}|M\rangle$ and terminates at $|G_\xi\rangle$ for the $B_\xi$ form, and at $|G_\eta\rangle$ for $B_\eta$ form. Here $|M\rangle = (a^+)^{N-1}|0\rangle$, and we use standard notation: $a^+$ and $a$ ($b^+$ and $b$) are the creation and annihilation operator for *A* (*B*) bosons, $|0\rangle = |0\rangle_A \otimes |0\rangle_B$ is the vacuum state of the whole boson system and *N* is the total number of particles.

(*iii*) The pointer states $|G_\xi\rangle$ and $|G_\eta\rangle$, and excitations around them belong to disconnected manifolds so that, following a measurement, our detector ends up in a mixed (as opposed to a pure, coherent superposition) state involving different vacua [10]. Thus, our model relates to approaches linking measurement events with spontaneous symmetry breaking [11,12,13,14,15].

We consider the hamiltonian $H = H_D + H_R + V'$ where $H_D$ and $H_R$ are, respectively, the hamiltonian of the detector and the thermal bath (at temperature $T$) and $V'$ describes the interaction between the bosons and the bath. $H_D = H_0 + V_{AB} + V_{BB}$ where $H_0 = \hbar\Omega(n_A + n_B^\xi + n_B^\eta)$ is the non-interacting term, and

$$V_{AB} = -\frac{\Lambda_\xi}{N}(b_\xi^+ a^+ b_\xi b_\xi + b_\xi^+ b_\xi^+ a\, b_\xi) - \frac{\Lambda_\eta}{N}(b_\eta^+ a^+ b_\eta b_\eta + b_\eta^+ b_\eta^+ a\, b_\eta)\ , \qquad (2)$$

and $V_{BB} = (\Xi/N) n_B^\xi n_B^\eta$ describe boson-boson interactions. $n_A = a^+ a$ and $n_B^{\xi,\eta} = b_{\xi,\eta}^+ b_{\xi,\eta}$ are number operators, and $\Xi \gg \Lambda_\xi, \Lambda_\eta > 0$ are coupling constants. Note that $H_D$ commutes with $N = n_A + n_B^\xi + n_B^\eta$. The two main diagrams for the non-repulsive component of the boson-boson interaction are shown in Fig. 2. Since $H_D|0_B\rangle = 0$, a state with an arbitrary number of $A$, but no $B$ bosons, is an exact eigenstate of $H_D$ with $E = N\hbar\Omega$. This applies in particular to the 'ready'-state of the detector $|M\rangle$.

In the following, we focus our attention on eigenstates $|\Psi\rangle$ for which $n_B^\xi n_B^\eta |\Psi\rangle \equiv 0$ (the assumption $\Xi \gg \Lambda_\xi, \Lambda_\eta$ effectively removes from the problem states with a macroscopic population of both types of bosons). Thus, the device eigenenergies

can be calculated by considering separately $B_\xi$ and $B_\eta$ states, and solving Schrödinger's equation for the reduced hamiltonian

$$\tilde{H} = \hbar\Omega(a^+ a + b^+ b) - \frac{\Lambda}{N}(b^+ a^+ b\, b + b^+ b^+ a\, b) \qquad . \qquad (3)$$

We search for solutions of the form

$$|\Psi\rangle = \sum_{s=1}^{N} C_s (a^+)^{N-s} (b^+)^s |0\rangle \qquad . \qquad (4)$$

Notice that, within this particular subspace, the number of eigenstates is equal to the number of particles and, therefore, that the entropy of the detector is always equal to zero. Replacing $C_s = \Theta_s / s(s-1)!!(N-s)!!$, the coefficients satisfy

$$\Lambda \frac{s(s-1)!!(N-s)!!}{N(s-2)!!(N-1-s)!!}(\Theta_{s+1} + \Theta_{s-1}) + (E - N\hbar\Omega)\Theta_s = 0 \quad . \qquad (5)$$

It follows that if $E = N\hbar\Omega + \Delta$ is an eigenenergy so is $E = N\hbar\Omega - \Delta$, and that the corresponding states are related by a change of sign of the even or odd coefficients. Also, considering the boundary conditions $\Theta_2/\Theta_1 = \Theta_{N-1}/\Theta_N = \frac{(N\hbar\Omega - E)(N-2)!!}{\Lambda(N-1)!!}$, we have that $E = N\hbar\Omega$ with $\Theta_s \equiv 0$ (even $s$) and $\Theta_s/\Theta_{s-2} = -1$ (odd $s$) is an exact solution for odd $N$. The eigenenergies obtained from direct numerical diagonalization of (5) are shown in Fig. 3. An analytical expression for the spectrum can be derived as follows. For $s \gg 1$ and even $N \gg s$, we approximate $\frac{s(s-1)!!(N-s)!!}{(s-2)!!(N-1-s)!!} \approx s^{3/2}(N-s)^{1/2} \equiv N^2 x^{3/2}(1-x)^{1/2}$ where $x = s/N$ is the concentration of $B$-particles. Provided $E < N\hbar\Omega$ so that $\Theta_s$ is smooth, we take the continuum limit of (5) [16]:

$$-\frac{1}{N}\frac{d^2\Theta}{dx^2} + NU(x)\Theta = 0 \qquad (6)$$

with

$$U = \frac{|E/N - \hbar\Omega|/\Lambda}{x^{3/2}(1-x)^{1/2}} - 2 \quad . \tag{7}$$

Equation (6) is identical to Schrödinger's equation of a particle of mass $N$ and zero eigenenergy moving in a one-dimensional gravitational-like potential [17], which diverges at $x = 0, 1$ and has an absolute minimum at $x_C = 3/4$. Replacing $\Theta(x) = \sin[f(x)]$ and, using the WKB approximation, we get $df/dx \approx N\sqrt{-U(x)}$. For $N \to \infty$, the eigenergies can be obtained by imposing the boundary condition

$$f(x_2) - f(x_1) = \int_{x_1}^{x_2} N\sqrt{-U(x)}dx = k\pi \tag{8}$$

where $U(x_{1,2}) = 0$ and $k$ is an integer. This expression gives values which agree extremely well with the numerical results. It can also be shown from (8) that $|E/N - \hbar\Omega|/\Lambda \propto (k/N)^{3/2}$ for $|E/N - \hbar\Omega| \ll \Lambda$.

The fact that $U \geq U(x_C = 3/4)$ simplifies considerably the analysis of the low-energy properties of our model. In the vicinity of $x_C$, we expand $U$ to second order in $(x - x_C)$ and, thus, relate $\Theta(x)$ to the wavefunction of a simple harmonic oscillator. A straightforward calculation shows that its frequency is $\Omega' = (3/\sqrt{2})\Lambda/\hbar$. Using the oscillator analogy, we can also find the ground-state energy, $E_G$, and the coefficients $\Theta_G(x)$ defining the ground-state wavefunction. Expanding in powers of $1/N$ we get, to lowest order, $E_G/N - \hbar\Omega = -(3\sqrt{3}/8)\Lambda$ and

$$\Theta_G \approx \exp{-(x - x_C)^2/\sigma^2} \tag{9}$$

where $\sigma \propto 1/N^{1/2}$. This indicates that ¾ of the $A$ bosons transform into $B$ particles while going from $|M\rangle$ to $|G\rangle$ [18] and that ground-state fluctuations in the number of particles are negligible. With these results in mind, it is convenient for $N \to \infty$ to lift the restriction that the total number of particles be conserved and use, instead of (4), an expression involving coherent states. Let $|\alpha_A\rangle$ and $|\alpha_B\rangle$ be the (coherent) eigenstates of the corresponding annihilation operators with eigenvalues $\alpha_A, \alpha_B > 0$. The *Ansatz*

$$|G'\rangle = |\alpha_A\rangle \otimes |\alpha_B\rangle \quad (11)$$

with $\alpha_A^2 = \langle \alpha_A | a^+ a | \alpha_A \rangle = N_A$ and $\alpha_B^2 = \langle \alpha_B | b^+ b | \alpha_B \rangle = N_B$ ($N_A + N_B = N$) gives, by minimizing $\langle G' | \tilde{H} - N\hbar\Omega | G' \rangle$, $E_{G'} = E_G$ and, as before, $N_A = \frac{1}{4}N$, $N_B = \frac{3}{4}N$ [19]. Hence, the ground state can be written simply as a product of *A* and *B* states showing broken global gauge symmetry. The correspondence with the harmonic oscillator problem indicates that the low-lying excitations are new quasiparticles (bosons) of frequency $\Omega'$. Using $(\alpha_B a - \alpha_A b)|G'\rangle \equiv 0$, it can be shown after some algebra that $(\alpha_B a^+ - \alpha_A b^+)|G'\rangle$ is a good approximation to the state with one quasiparticle for a system of $N + 1$ bosons.

We have so far limited our discussion to the Hamiltonian of the device. However, it is clear that, starting at $|M\rangle$, the pointer states $|G_\xi\rangle$ and $|G_\eta\rangle$ can only be reached with assistance of the thermal bath, through $V'$. Assuming that $V'$ conserves the number of both $B_\xi$ and $B_\eta$ and, therefore, that it does not mix the corresponding Hilbert spaces, it is also apparent that the initial states $b_\xi^+ |M\rangle$ and $b_\eta^+ |M\rangle$ will evolve, respectively, into superpositions of low-lying levels associated with $|G_\xi\rangle$ and $|G_\eta\rangle$ (note that the number of quasiparticles of frequency $\Omega'$ must be finite, *i. e.*, non-macroscopic, when the apparatus attains thermal equilibrium). Accordingly, hybrid microscopic superpositions of the form $(c_\xi b_\xi^+ + c_\eta b_\eta^+)|M\rangle$ will evolve not into a pure state, but a mixed one (diagonal density ma-

trix). Hence, decoherence is an intrinsic component of our detector's operation. Provided $V'$ is sufficiently weak, these considerations also apply to the more interesting case where the bath can induce transitions and, thus, tunneling between the $B_\xi$ and $B_\eta$ subspaces. $B_\xi$ - $B_\eta$ mixing is expected to be important only in the earliest stages of the measuring process. In the thermodynamic limit, as the detector moves down the ladder of states, its state will become more susceptible to the macroscopic character of the respective populations (or, alternatively, to symmetry breaking). Eventually, the device will reach a mixed state in which the bath will no longer be able to induce tunneling between the $B_\xi$ and $B_\eta$ manifolds. Within this context, the occurrence of Schrödinger cat states is evidently related to the problem of superpositions of spontaneously-broken or, simply, inequivalent vacuum states, and that of wavefunction collapse to the dynamics of broken symmetry in a phase transformation. It remains to be seen whether a generic coupling to the bath will lead to true collapse of the wavefunction (by destroying the microscopic superpositions before they become macroscopic Schrödinger cat states) or that coherent superpositions will always evolve into mixed states, as in the 'decoherence' scenario.

In summary, we have described a fully quantum mechanical model which mimics the behavior of an avalanche detector and thereby reveals an unambiguous connection between the Schrödinger cat (or, alternatively, the wavefunction collapse) problem and the question of macroscopic superpositions involving multiple vacua. Because of its simplicity, our model holds promise for time-domain studies of spontaneous symmetry breaking and for elucidating the role of the environment in a quantum measurement.

FIGURE CAPTIONS

FIG. 1 (color online) – Schematic diagram of the quantum detector. The initial state $|M\rangle$ contains $N$-1 bosons of the $A$–type. A single $B_\xi$ ($B_\eta$) particle sets off a bath-assisted transition into the final state $|G_\xi\rangle$ ($|G_\eta\rangle$) where ¾ of the particles are of the $B$-type.

FIG. 2 (color online) – The two scattering diagrams associated with $V_{AB}$.

FIG. 3 (color online) – Energy per particle calculated for $N = 128$. Modes are labeled with the quantum number $k$, with $0 < k \leq N/2$, following the highest-to-lowest (lowest-to-highest) order for $E < N\hbar\Omega$ ($E > N\hbar\Omega$).

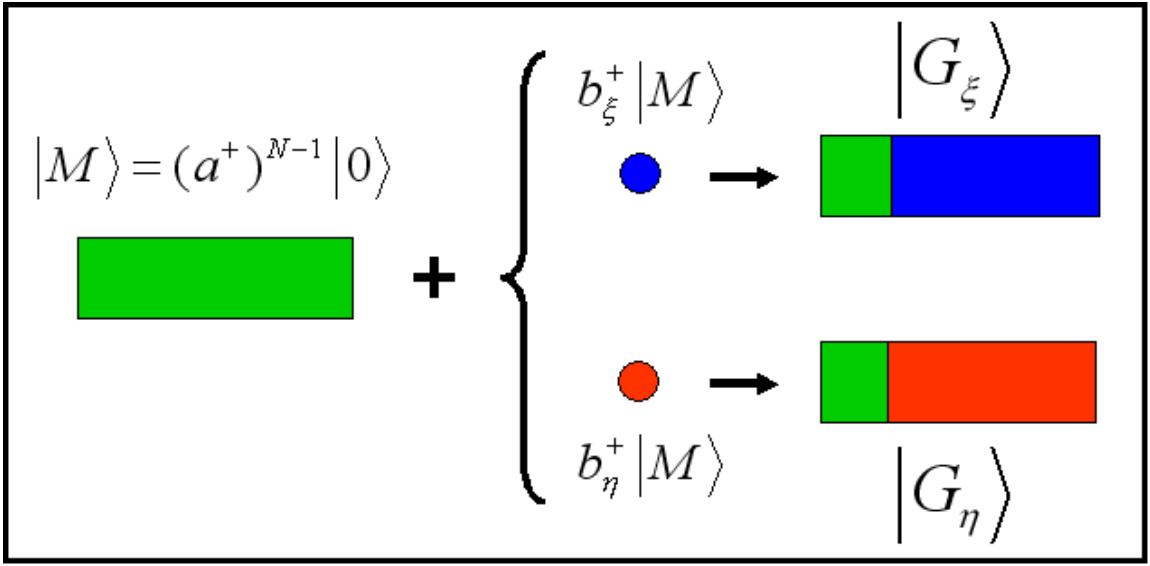

FIGURE 1

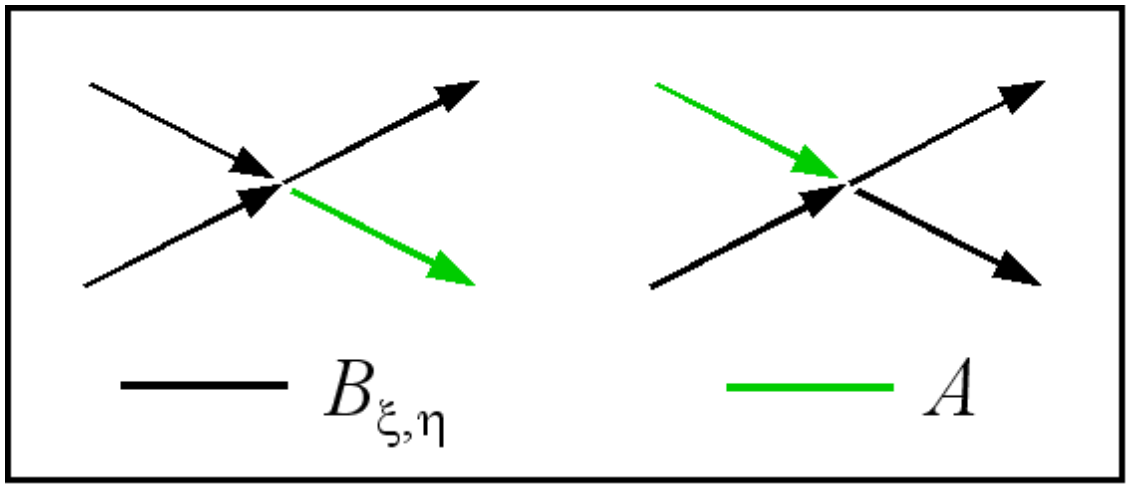

FIGURE 2

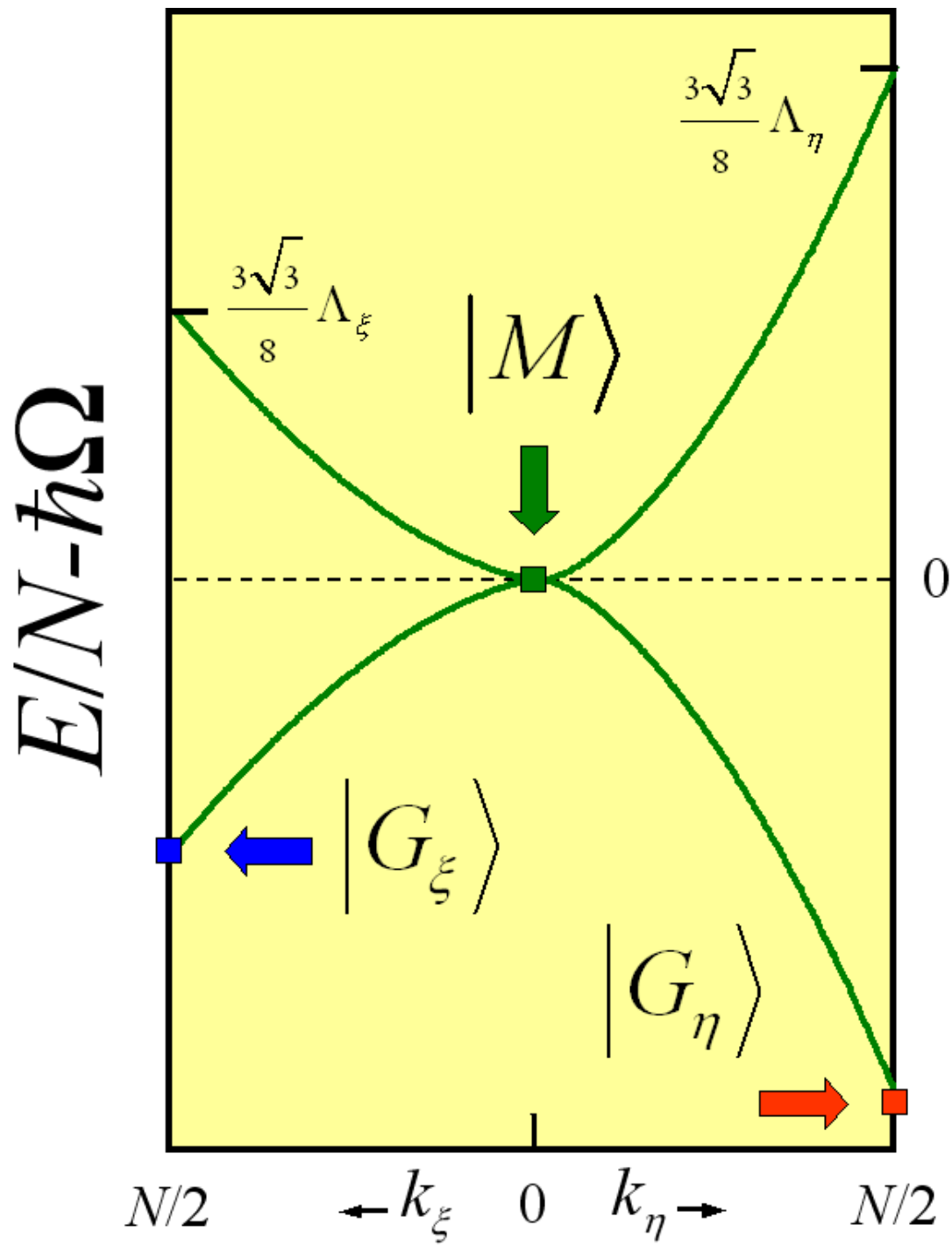

FIGURE 3